\documentclass[twocolumn,aps,pra,longbibliography]{revtex4-2}

\usepackage{graphicx}
\usepackage{amsmath,amssymb}
\usepackage{xcolor}

\usepackage[colorlinks=true,
            linkcolor=blue,
            citecolor=blue,
            urlcolor=blue]{hyperref}
\begin{document}
\title{Frequency upconversion of infrared signals via molecular cavity optomechanical systems with gain}

\author{Shu-Xian Quan}
\author{Fen Zou}
\email{zoufen@hainanu.edu.cn}
\author{Yong Li}
\email{yongli@hainanu.edu.cn}
\affiliation{Center for Theoretical Physics, School of Physics and Optoelectronic Engineering, Hainan University, Haikou 570228, China}

\date{\today}
\begin{abstract}
 Molecular cavity optomechanical systems have recently emerged as a promising platform for enhancing infrared detection sensitivity, owing to their ability to up-convert low-frequency infrared (IR) photons to visible frequency range.  Generally, under red-detuned pumping in such systems, the ideal conversion efficiency of the IR signal  approaches 1. To overcome this efficiency constraint, we propose a scheme that incorporates gain into the infrared cavity of a molecular cavity  optomechanical system comprising two cavities and an ensemble of $N$ molecules. The upconversion process, which relies on IR absorption and Raman scattering associated with specific vibrational modes, is significantly amplified by the incorporation of gain under the red-detuned conditions. Moreover, our analysis demonstrates that the added noise is maintained near 0.5.
\end{abstract}
\maketitle
\section{introduction}
The infrared (IR) spectrum spanning from the mid-infrared (MIR) to far-infrared (FIR) wavelength bands effectively provides information on molecular arrangements, optical responses, and thermal processes that are inaccessible in other regions of the electromagnetic spectrum. This unique property has led to the widespread application of IR spectroscopy in various fields, including thermal imaging~\cite{tonouchi2007cutting}, astronomical surveys~\cite{10.1117/1.JATIS.6.4.041503}, and remote sensing~\cite{ng2009remote}. The rapid development of IR spectroscopy has brought a pressing demand for IR detectors characterized by high sensitivity, low noise, and cost-effectiveness. Despite advances in new detectors tailored for the IR range ~\cite{10.1063/1.4985060,Ariyoshi_2016}, the high thermal noise inherent to IR radiation remains a major bottleneck in achieving high-precise measurements. 
Frequency upconversion techniques offer a solution by converting FIR or MIR signals into visible (VIS) or near-infrared (NIR) ranges that are readily available for well-established and low-noise detection devices. 

The conventional method to realize frequency upconversion is based on three-wave mixing processes in bulk nonlinear optics~\cite{Temporao:06,Tseng:18}, which requires precise phase matching and high peak power. With the advancement of quantum technologies, cavity optomechanical systems, as hybrid platforms coupling optical field to mechanical motion, are proposed to facilitate coherent wavelength conversion mediated by mechanical resonators. Such a process enables incident coherent signals to achieve high-fidelity conversion~\cite{hill2012coherent,PhysRevA.97.043818,Li:17,PhysRevA.96.053853,PhysRevA.93.023827,doi:10.1126/science.1228370,palomaki2013coherent,PhysRevLett.108.153603,bochmann2013nanomechanical,andrews2014bidirectional,PhysRevLett.113.023604,PhysRevLett.112.133904,forsch2020microwave,PhysRevLett.120.023601}. 

Molecular cavity optomechanical systems ~\cite{PhysRevA.111.043507,yu2025strongmoleculelightentanglementmolecular,Schmidt_2024,MoradiKalardeCiccarelloS,PhysRevResearch.7.013161,PENG2025116473,Tang2026} extend the standard cavity optomechanical framework by substituting macroscopic mechanical resonators with intrinsic molecular vibrational modes. Specifically, the coupling originates from a shift of the plasmon resonance frequency led by the change in polarizability of the vibrational molecule~\cite{roelli2016molecular}.  Compared with cavity optomechanical systems, molecular cavity optomechanical systems possess the extremely large optomechanical coupling strength  and high-frequency molecular vibrational modes. Owing to these advantages, molecular cavity optomechanical systems have emerged as a promising platform to achieve frequency conversion. In such a system driven by a pump field, the low-frequency signal incident on an IR cavity can be upconverted to the Stokes or anti-Stokes sideband of the pump field in the VIS domain via IR absorption that excites molecular vibrations, which in turn give rise to the Raman scattering mediated by the vibrational modulation of molecular polarizability~\cite{PhysRevX.10.031057,xomalis2021detecting,doi:10.1126/science.abk3106}.

Benefiting from the intrinsically low thermal occupancy of high-frequency molecular vibrational modes, molecular cavity optomechanical systems allow high-fidelity conversion even at ambient temperature. Moreover, the extremely large coupling strength resulting from the strong localized optical field in plasmonic cavities facilitates  efficient frequency conversion. In the case of a red-detuned pump field, the frequency upconversion is dominated by the anti-Stokes process. The maximum conversion efficiency from the IR to the VIS range generally approaches unity. To obtain a higher frequency conversion, Zou \textit{et al.}~\cite{zou2024amplifying,zou2026frequencyupconversioninfraredsignals} employed a blue-detuned pump field to drive the molecular cavity optomechanical system and theoretically demonstrated that the intensity of the IR signal upconverted to the VIS range can be enhanced by a factor of 1000 in the first Stokes sideband of the pump field.

Motivated by the molecular cavity optomechanical systems employing red and blue-detuned pumps~\cite{berinyuy2025enhancing,zou2024amplifying,roelli2016molecular}, we propose an alternative approach that leverages a gain-assisted molecular cavity optomechanical system to achieve the enhancement of the  upconversion efficiency of IR signal. Our approach is implemented under a red-detuned pump field, where the gain involved is incorporated into the IR cavity. In this paper, we investigate the effect of the gain incorporated in the IR cavity on the frequency upconversion efficiency and the associated added noise. To develop further insight, we explore the dependence of the conversion efficiency on other key system parameters, such as the decay rates of the molecular vibrational mode and the VIS cavity. Our results demonstrate that gain-assisted molecular cavity optomechanical systems can not only efficiently enhance the conversion efficiency of the IR signal but also keep the associated added noise at a low and controllable level. These findings offer a theoretical foundation for improving frequency upconversion technology.

\section{molecular cavity optomechanical systems}\label{a}

\begin{figure}
\centering{}\includegraphics[width=9cm,totalheight=5.5cm]{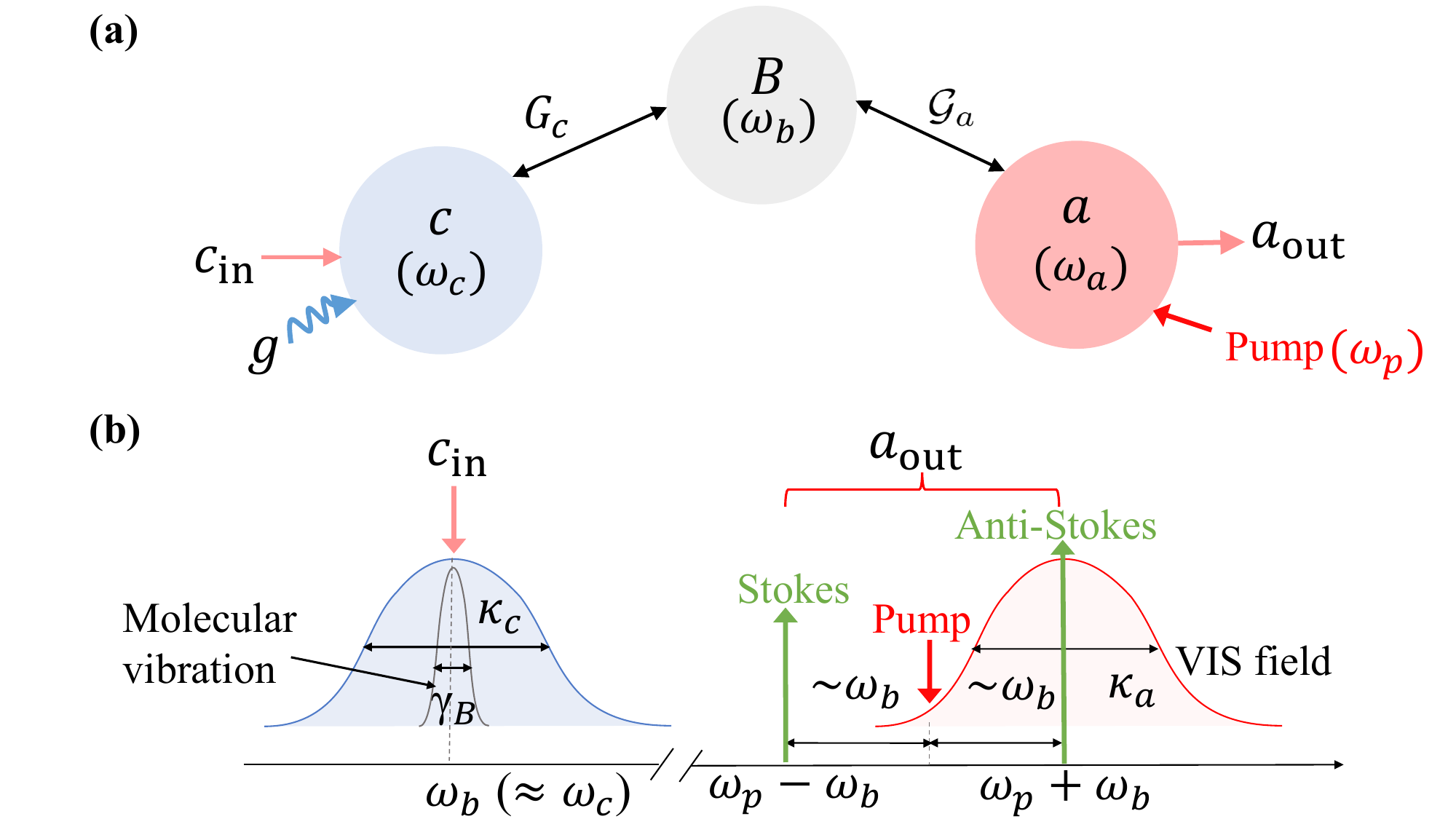}\caption{(a) Schematic diagram of the molecular cavity optomechanical system. Here, the collective mode $B$ composed of $N$ molecules (vibrational frequency $\omega_{b}$) is coupled to both the VIS mode (frequency $\omega_{a}$, decay rate $\kappa_{a}$) via optomechanical interaction and the IR mode (frequency $\omega_{c}$, decay rate $\kappa_{c}$, and gain rate $g$) via bilinear interaction.
The VIS cavity is driven by a pump field with frequency $\omega_{p}$
and amplitude $\varepsilon_{p}.$ (b) Schematic illustration diagram of the IR to VIS upconversion in a molecular optomechanical system driven by the red-detuned pump field, where $c_\text{in}$  includes the incident IR signal, and the cavity frequency is considered to be close to the molecular vibrational frequency ($\omega_b\approx\omega_c$).} \label{fig:1}
\end{figure}

The molecular cavity optomechanical system under consideration, schematically described in Fig.~\ref{fig:1}(a), consists of $N$ identical molecules coupled to both a cavity of IR mode and a plasmonic one of VIS mode. Here, we consider gain is incorporated into the IR cavity. A strong pump field with frequency $\omega_{p}$ and amplitude $\varepsilon_{p}$ is applied to drive the VIS cavity. Under low excitation, the vibrational mode of each molecule can be approximated as a harmonic oscillator with frequency $\omega_{b}$~\cite{roelli2016molecular}. The Hamiltonian of this system is given as ($\hbar=1$)~\cite{zou2024amplifying}
\begin{align}
H & =\omega_{a}a^{\dagger}a+\omega_{c}c^{\dagger}c+\sum_{j=1}^{N}\omega_{b}b_{j}^{\dagger}b_{j}+\sum_{j=1}^{N}g_{a}a^{\dagger}a(b_{j}^{\dagger}+b_{j})\nonumber \\
 &\quad +\sum_{j=1}^{N}g_{c}(c^{\dagger}+c)(b_{j}^{\dagger}+b_{j})+(i\varepsilon_{p}e^{-i\omega_{p}t}a^{\dagger}+\textrm{H.c.}).\label{eq:1}
\end{align}
Here, $a$ and $c$ are respectively the annihilation operators of the VIS mode with frequency $\omega_{a}$ and the IR mode with frequency $\omega_{c}$, and $b_{j}$ is the annihilation operator for the vibrational mode of the $j$-th molecule. The VIS and  IR modes are optomechanically and bilinearly coupled to the vibrational mode of the  molecule, whose coupling strengths are represented by $g_{a}$ and  $g_{c}$, respectively~\cite{schmidt2016quantum,benz2016single,roelli2016molecular,lombardi2018pulsed,zhang2020optomechanical,esteban2022molecular,xu2022phononic,shalabney2015coherent,pannir2022driving}. The total contribution of all molecules can be effectively represented by a collective vibrational operators $B=\frac{1}{\sqrt{N}}\sum_{j=1}^{N}b_{j}$ that satisfies the commutation relation $\left[B,B^{\dagger}\right]=1$. In the interaction picture with respect to $\omega_{p}a^{\dagger}a$, the system is described by the Hamiltonian
\begin{eqnarray}
H_{I} & = & \Delta_{0}a^{\dagger}a+\omega_{c}c^{\dagger}c+\omega_{b}B^{\dagger}B_{}+G_{a}a^{\dagger}a(B_{}^{\dagger}+B)\nonumber \\
 &  &+G_{c}(c^{\dagger}+c)(B_{}^{\dagger}+B_{})+(i\varepsilon_{p}a^{\dagger}+\textrm{H.c.}),\label{eq:2}
\end{eqnarray} where $\Delta_{0}=\omega_{a}-\omega_{p}$ is the detuning between the VIS mode and the pump field, and $G_{a}=g_{a}\sqrt{N}$ and $G_{c}=g_{c}\sqrt{N}$ represent the collective optomechanical and bilinear coupling strengths, respectively.

Under the assumption of two single-sided cavities with negligible intrinsic losses, the quantum Langevin equations (QLEs) are obtained as:
\begin{eqnarray}
\frac{da}{dt} & = & -(i\Delta_{0}+\kappa_{a})a-iG_{a}a(B^{\dagger}+B)+\varepsilon_{p}+\sqrt{2\kappa_{a}}a_{\textrm{in}},\nonumber \\
\frac{dc}{dt} & = & -(i\omega_{c}+\zeta)c-iG_{c}(B^{\dagger}+B)+\sqrt{2\kappa_{c}}c_{\textrm{in}}+\sqrt{2g}c^{(g)}_{\textrm{in}},\nonumber \\
\frac{dB}{dt} & = & -(i\omega_{b}+\gamma_{B})B-iG_{a}a^{\dagger}a-iG_{c}(c^{\dagger}+c)\nonumber \\
 &  & +\sqrt{2\gamma_{B}}B_{\textrm{in}}.\label{eq:3}
\end{eqnarray}
Here, the effective dissipation rate $\zeta=\kappa_{c}-g$ is defined as the difference between the decay rate $\kappa_{c}$  and the incorporated gain rate $g$ of the IR cavity; the decay rates of the VIS mode and the collective molecular vibrational mode are $\kappa_{a}$ and $\gamma_{B}$, respectively.  Each input operator  ${o}_{\mathrm{in}}$ ($o=a,B,c$) comprises not only the noise associated with the thermal bath but also possible externally incident signal components. Under the condition $\hbar\omega_{a,c,b}/k_{B}T\gg1$ that holds for most experimentally relevant parameters  of molecular cavity optomechanical systems, the thermal occupations of all modes are negligible, so that the related noise can be treated as vacuum noise. Here, $k_{B}$ denotes the Boltzmann constant and $T$ is the temperature of the environment. In the absence of external signal, the mean values and correlation functions of these input operators have $\left\langle o_\text{in}(t)\right\rangle=0$ and $\left\langle o_\text{in}(t)o_\text{in}^\dagger(t')\right\rangle=\delta(t-t')$~\cite{PhysRevA.94.031802}. The gain operator $c^{(g)}_{\text{in}}$ likewise has a vanishing mean value $\left\langle c^{(g)}_{\text{in}}(t)\right\rangle=0$ and nonzero correlation $\left\langle c^{(g)\dagger}_{\text{in}}(t)c^{(g)}_{\text{in}}(t')\right\rangle=\delta(t-t')$~\cite{liu2017energy,PhysRevA.85.031802}. Hence, the steady-state solutions of the system are expressed as
\begin{eqnarray}
\left\langle a\right\rangle _{\textrm{ss}} & = & \frac{\varepsilon_{p}}{i\Delta+\kappa_{a}},\nonumber \\
\left\langle c\right\rangle _{\textrm{ss}} & = & \frac{-iG_{c}}{i\omega_{c}+\zeta}(\left\langle B\right\rangle_{\textrm{ss}} ^{*}+\left\langle B\right\rangle_{\textrm{ss}} ),\nonumber \\
\left\langle B\right\rangle _{\textrm{ss}} & = & -\frac{iG_{a}\left|\left\langle a\right\rangle_\text{ss} \right|^{2}+iG_{c}(\left\langle c\right\rangle_\text{ss} ^{*}+\left\langle c\right\rangle_\text{ss} )}{i\omega_{b}+\gamma_{B}},
\end{eqnarray}
where $\Delta=\Delta_{0}+G_{a}(\left\langle B\right\rangle_\text{ss} ^{*}+\left\langle B\right\rangle _\text{ss})$
denotes the effective detuning that includes the frequency shift induced by the collective optomechanical interaction. 

We decompose the operator $o$ into the mean steady-state value and the quantum fluctuation term, namely $o=\left\langle o\right\rangle _\text{ss}+\delta o$~\cite{weis2010optomechanically}. Substituting this decomposition into QLEs~(\ref{eq:3}) and retaining only the first-order terms in the quantum fluctuation operator $ \delta o$, we can derive the linearized QLEs as (replacing $\delta o$ with $o$ hereafter)
\begin{align}
\frac{da}{dt} & =  -(i\Delta+\kappa_{a})a-i\mathcal{G}_{a}( B^{\dagger}+B)+\sqrt{2\kappa_{a}}a_{\textrm{in}},\nonumber \\
\frac{d c}{dt} & = -(i\omega_{c}+\zeta) c-iG_{c}(B^{\dagger}+B)+\sqrt{2\kappa_{c}}c_{\textrm{in}}\nonumber \\
 & \quad  +\sqrt{2g}c^{(g)}_{\text{in}},\nonumber \\
\frac{d B}{dt} & = -(i\omega_{b}+\gamma_{B}) B- i\mathcal{G}_{a}\text{(} a+a^{\dagger})\nonumber\\
  &\quad-iG_{c}(c^{\dagger}+c)+\sqrt{2\gamma_{B}}B_{\textrm{in}}.\label{eq:5}
\end{align}
For simplicity but without loss of generality, the enhanced collective optomechanical coupling strength $\mathcal{G}_{a}=G_{a}\left\langle a\right\rangle _{\textrm{ss}}$ has been taken to be a  positive real number.  By defining two fluctuation vectors $V=(a$, $ c$, $ B$, $a^{\dagger}$, $ c^{\dagger}$, $ B^{\dagger})^{\text{T}}$ and $V_{\textrm{in}}=( a_{\textrm{in}}$,
$ c_{\textrm{in}}$, $ B_{\textrm{in}}$, $ c_{\textrm{in}}^{(g)}$,
$ a_{\textrm{in}}^{\dagger}$, $c_{\textrm{in}}^{\dagger}$, $B_{\textrm{in}}^{\dagger}$, $c^{(g)\dagger}_{\text{in}}$$)^{\textrm{T}}$, Eq.~(\ref{eq:5}) is rewritten in the matrix form as
\begin{eqnarray}
\frac{dV}{dt} & = & -MV+\Gamma V_{\textrm{in}}.\label{eq:6}
\end{eqnarray}
Here, $M=\left(\begin{array}{cc}
P & Q\\
Q^{*} & P^{*}
\end{array}\right)$ is the coefficient matrix and $\Gamma=\left(\begin{array}{cc}
\chi & 0\\
0 & \chi
\end{array}\right)$ denotes the damping matrix, where 
\begin{eqnarray}
P & = & \left(\begin{array}{ccc}
i\Delta+\kappa_{a} & 0 & i\mathcal{G}_{a}\\
0 & i\omega_{c}+\zeta & iG_{c}\\
i\mathcal{G}_{a} & iG_{c} & i\omega_{b}+\gamma_{B}
\end{array}\right),\label{eq:7}
\end{eqnarray}
\begin{eqnarray}
Q & = & \left(\begin{array}{ccc}
0 & 0 & i\mathcal{G}_{a}\\
0 & 0 & iG_{c}\\
i\mathcal{G}_{a} & iG_{c} & 0
\end{array}\right),\label{eq:8}
\end{eqnarray}
and

\begin{eqnarray}
\chi & = & \left(\begin{array}{cccc}
\sqrt{2\kappa_{a}} & 0 & 0 & 0\\
0 & \sqrt{2\kappa_{c}} & 0 & \sqrt{2g}\\
0 & 0 & \sqrt{2\gamma_{B}} & 0
\end{array}\right).\label{eq:9}
\end{eqnarray}
According to the Routh-Hurwitz criterion~\cite{dejesus1987routh}, the stability of the system requires that the real parts of all the eigenvalues of the coefficient matrix $M$ are positive. 

To evaluate the conversion efficiency from the IR signal to the VIS range, we analyze the output spectrum of the VIS mode in the frequency domain. The relevant operators in the frequency domain are expressed through the Fourier transform as
\begin{eqnarray}
\widetilde{o}[\omega] & = & \frac{1}{\sqrt{2\pi}}\int_{-\infty}^{\infty}o(t)e^{i\omega t}dt,\nonumber \\
\widetilde{o^{\dagger}}[\omega] & = & \frac{1}{\sqrt{2\pi}}\int_{-\infty}^{\infty}o^{\dagger}(t)e^{i\omega t}dt,\label{eq:10}
\end{eqnarray}
which obeys $\widetilde{o}[\omega]^{\dagger}=\widetilde{o^{\dagger}}[-\omega]$ ($o=a, B, c$). By applying this transformation to Eq.~(\ref{eq:6}), the fluctuation vector  $\widetilde{V}[\omega]$ in the frequency domain is given by
\begin{equation}
\widetilde{V}[\omega]=(M-i\omega I_{6})^{-1}\Gamma \widetilde{V}_{\text{in}}[\omega],\label{eq:11}
\end{equation}
where $I_{6}$ is the 6$\times6$ identity matrix. According to the input-output relation $\widetilde{V}_{\textrm{in}}[\omega]+\widetilde{V}_{\textrm{out}}[\omega]=\Gamma^{\textrm{T}}\widetilde{V}[\omega]$~\cite{gardiner1985input}, the output field vector in the frequency domain satisfies 
\begin{equation}
\widetilde{V}_{\textrm{out}}[\omega]=U(\omega)\widetilde{V}_{\textrm{in}}[\omega]\label{eq:12}
\end{equation}
with the scattering matrix $U(\omega)$ given by
\begin{equation}
U(\omega)=\Gamma^{\textrm{T}}(M-i\omega I_{6})^{-1}\Gamma-I_{8}.\label{eq:13}
\end{equation}
Consequently, the output field operator of the VIS cavity is expressed as
\begin{align}
\widetilde{a}_{\textrm{out}}[\omega] & =  U_{11}(\omega)\widetilde{a}_{\textrm{in}}[\omega]+U_{12}(\omega)\widetilde{c}_{\textrm{in}}[\omega]+U_{13}(\omega)\widetilde{B}_{\textrm{in}}[\omega]\nonumber\\
  &\quad+U_{14}(\omega)\widetilde{c}^{(g)}_{\text{in}}[\omega]+U_{15}(\omega)\widetilde{a_{\textrm{in}}^{\dagger}}[\omega]+U_{16}(\omega)\widetilde{c_{\textrm{in}}^{\dagger}}[\omega]\nonumber\\
  &\quad+U_{17}(\omega)\widetilde{B_{\textrm{in}}^{\dagger}}[\omega]+U_{18}(\omega)\widetilde{c^{(g)\dagger}_{\text{in}}}[\omega].\label{eq:14}
 \end{align}
The matrix element $U_{ij}$ represents the transmission amplitude from the input mode
$j$ ($j$-th element in the input field vector) to the output mode $i$ ($i$-th
element in the output field vector). Due to their cumbersome length, the explicit expressions are therefore not included. In the presence of the possible incident signal, the correlation functions of the input operator $\widetilde{o}_{\textrm{in}}[\omega]$ are given by~\cite{Tth2017,PhysRevA.91.053854}
\begin{align}
\left\langle\widetilde{o_{\textrm{in}}^{\dagger}}[\omega]\widetilde{o}_{\textrm{in}}[\omega']\right\rangle &=  S_{o, \text{in}}(\omega)\delta(\omega+\omega')\nonumber \\
\left\langle\widetilde{o}_{\textrm{in}}[\omega']\widetilde{o_{\textrm{in}}^{\dagger}}[\omega']\right\rangle&=\left[S_{o, \text{in}}(\omega)+1\right]\delta(\omega+\omega') ,\label{eq:50}
 \end{align}
where $S_{o, \text{in}}(\omega)$ originates from the signals incident on mode $o$, and the other term 1 is introduced by the vacuum noise. Furthermore, analogous definition is extended to the correlation function of the input operator associated with the gain in the IR cavity, following~\cite{jiang2018directional,PhysRevA.85.021801}
\begin{align}
\left\langle\widetilde{c_{\textrm{in}}^{(g)\dagger}}[\omega]\widetilde{c}_{\textrm{in}}^{(g)}[\omega']\right\rangle &= \delta(\omega+\omega')\nonumber \\
\left\langle\widetilde{c}_{\textrm{in}}^{(g)}[\omega']\widetilde{c_{\textrm{in}}^{(g)\dagger}}[\omega']\right\rangle&=0 .\label{eq:51}
 \end{align}The output spectrum of the VIS cavity is determined by~\cite{jiang2018directional,PhysRevLett.112.133904}
\begin{align}
\bar{S}_{a,\textrm{out}}(\omega) & =\frac{1}{2} \int\langle\widetilde{a_{\textrm{out}}^{\dagger}}[\omega]\widetilde{a}_{\textrm{out}}[\omega']+\widetilde{a}_\textrm{out}[\omega']\widetilde{a^\dagger_{\textrm{out}}}[\omega]\rangle d\omega'. \label{eq:17}
\end{align}Based on Eqs.~(\ref{eq:14}-\ref{eq:17}), one further gets
\begin{align}\bar{{S}}_{a,\textrm{out}}(\omega)=\sum_{o=a,c,B}T_{ao}(\omega)\left[S_{o,\textrm{in}}+\frac{1}{2}\right]+S_{c,\textrm{gain}}(\omega).\label{eq:18}
\end{align}The conversion efficiency from mode $o$ $(o=a, B, c)$ to the VIS mode and the noise spectrum induced by the gain are, respectively, given by
\begin{eqnarray}
T_{aa}(\omega)&=&\left|U_{11}(\omega)\right|^{2}+\left|U_{15}(\omega)\right|^{2},\nonumber \\
T_{ac}(\omega)&=&\left|U_{12}(\omega)\right|^{2}+\left|U_{16}(\omega)\right|^{2}, \nonumber \\
T_{aB}(\omega)&=&\left|U_{13}(\omega)\right|^{2}+\left|U_{17}(\omega)\right|^{2},\label{eq:19}
\end{eqnarray}
and 
\begin{eqnarray}
S_{c,\textrm{gain}}(\omega)&=&\frac{\left|U_{14}(\omega)\right|^{2}+\left|U_{18}(\omega)\right|^{2}}{2}.\label{eq:20}
\end{eqnarray}

With an external signal incident only on the IR mode, i.e., $S_{a,\textrm{in}}(\omega)=S_{B,\textrm{in}}(\omega)=0$, the output spectrum of the VIS cavity is simplified as~\cite{hill2012coherent,PhysRevX.6.041024,RevModPhys.82.1155}
\begin{align}
\bar{S}_{a,\textrm{out}}(\omega) = T_{ac}(\omega)\left[S_{c,\textrm{in}}(\omega)+\frac{1}{2}\right]+S_{\textrm{add}}(\omega).\label{eq:23}
\end{align}Here, the added noise power $S_{\textrm{add}}(\omega)=S_{c,\textrm{gain}}(\omega)+S_{a,\textrm{vac}}(\omega)+S_{B,\textrm{vac}}(\omega)$ with $S_{a,\textrm{vac}}(\omega)=\left(\left|U_{11}(\omega)\right|^{2}+\left|U_{15}(\omega)\right|^{2}\right)/2$ and  $S_{B,\textrm{vac}}(\omega)=\left(\left|U_{13}(\omega)\right|^{2}+\left|U_{17}(\omega)\right|^{2}\right)/2$  represents the contribution of  the gain-induced noise ($S_{c,\text{gain}}$) and of the vacuum noise from  $a$ modes as well as $B$ mode ($S_{a,\textrm{vac}}$ and  $S_{B,\textrm{vac}}$) to the output spectrum.
\begin{figure}
\centering{}\includegraphics[width=8.8cm,totalheight=4.4cm]{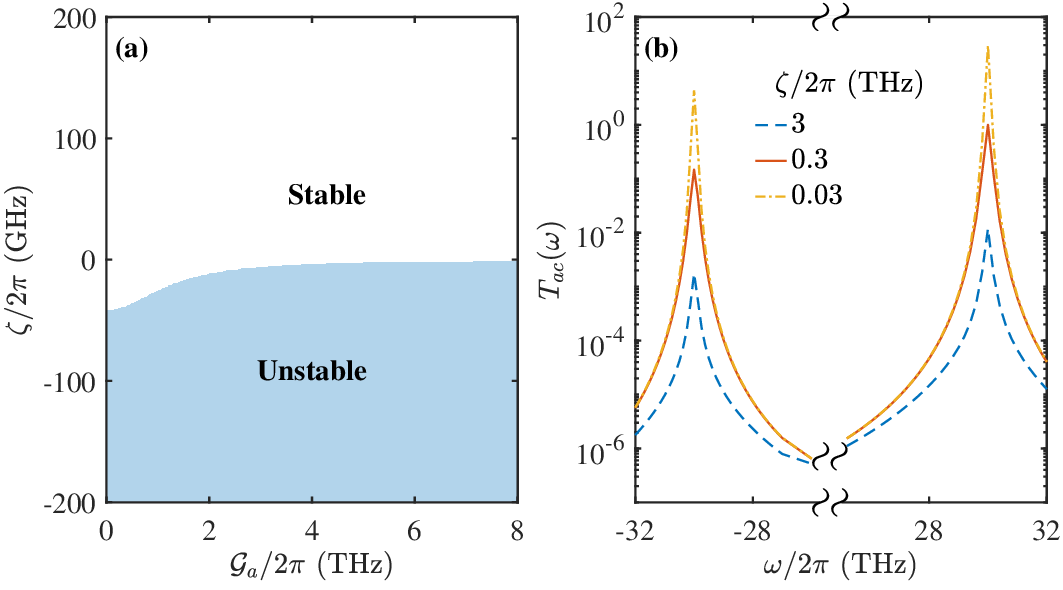}\caption{(a) The stability diagram associated with the enhanced optomechanical coupling strength $\mathcal{G}_{a}$ and the effective dissipation rate $\zeta$
of IR mode. (b) The conversion efficiency $T_{ac}(\omega)$ with respect
to the frequency $\omega$ of  incident signal for different effective
dissipation rate $\zeta$ with $\mathcal{G}_{a}/2\pi=800$ GHz. The parameters are $g_{c}/2\pi=0.05$ GHz, $\Delta/2\pi=\omega_{c}/2\pi=\omega_{b}/2\pi=30$ THz, $\kappa_{a}/2\pi=25$
THz, $\kappa_{c}/2\pi=3$ THz, $\gamma_{B}/2\pi=0.06$ THz, and $N=10^{6}$.\label{fig~2}}
\end{figure}

A prerequisite for the effective frequency conversion process is that the system is stable. Figure~\ref{fig~2}(a) shows the stability distribution in the parameter space $(\mathcal{G}_{a},\zeta)$, where the white (blue) area denotes the stable (unstable) region of the system. For the system driven by the red-detuned pump with $\Delta=\omega_b$, we plot the conversion efficiency $T_{ac}(\omega)$ as a function of frequency for different effective dissipation rates $\zeta$ in Fig.~\ref{fig~2}(b). The conversion efficiency from the IR signal to the VIS range exhibits distinct peaks at the Stokes and anti-Stokes sidebands, where the conversion efficiency at the anti-Stokes sideband is higher than that at the Stokes sideband. Without the gain (that means $\zeta=\kappa_c=2\pi\times3$ THz), the conversion efficiency stays below unity. By incorporating the gain into the IR cavity, the effective dissipation rate is reduced, leading to a significant enhancement in $T_{ac}(\omega)$. In the case where the gain $g$ exceeds the decay rate $\kappa_c$ ($\zeta<0$), the IR cavity enters a net-gain regime. As the effective dissipation rate $\zeta/2\pi$ is below $0.3$ THz, the peak value of the conversion efficiency exceeds 1. That means that this optomechanical system enables frequency upconversion of IR signals to the VIS regime while simultaneously amplifying the IR signals.

\section{red-detuned pump field with rotating-wave approximation}

In the molecular cavity optomechanical system driven by a red-detuned pump field with $\Delta\approx\omega_{c}\approx\omega_{b}\gg\left\{ G_{c},\mathcal{G}_{a}\right\}$, the fast-oscillating terms $\mathcal{G}_{a}(\delta a^{\dagger}\delta B^{\dagger}+\delta a\delta B)$ and $G_c(\delta c^{\dagger}\delta B^{\dagger}+\delta c\delta B)$ can be ignored under the rotating-wave approximation (RWA). As a result, the linear LEQs~(\ref{eq:5}) are modified as
\begin{eqnarray}
\frac{d a}{dt} & = & -(i\Delta+\kappa_{a}) a-i\mathcal{G}_{a}B+\sqrt{2\kappa_{a}}a_{\textrm{in}},\nonumber \\
\frac{d c}{dt} & = & -(i\omega_{c}+\zeta) c-iG_{c} B+\sqrt{2\kappa_{c}}c_{\textrm{in}} +\sqrt{2g}c_{\textrm{in},g}\nonumber \\
\frac{dB}{dt} & = & -(i\omega_{b}+\gamma_{B}) B-i\mathcal{G}_{a}a-iG_{c}c+\sqrt{2\gamma_{B}}B_{\textrm{in}}.\label{eq:24}
\end{eqnarray}
By transforming the relevant operators to the frequency domain and then constructing the vectors   $\widetilde{{V}}'[\omega]=(\widetilde{a}[\omega],\,\widetilde{c}[\omega],\,\widetilde{B}[\omega])^\text{T}$ and  $\widetilde{V}'_{\textrm{in}}[\omega]=(\widetilde{a}_{\textrm{in}}[\omega],\,\widetilde{c}_{\textrm{in}}[\omega],\,\widetilde{B}_{\textrm{in}}[\omega],\,\widetilde{c}^{(g)}_{\text{in}}[\omega])^\text{T}$, Eq.~(\ref{eq:24}) is recast into the matrix form as
\begin{eqnarray}
\widetilde{V}'[\omega] & = & (P-i\omega I_3)^{-1}\chi\widetilde{V}'_{\textrm{in}}[\omega].\label{eq:25}
\end{eqnarray}
Based on the input-output relation mentioned in  Sec.~\ref{a}, the output field vector in the frequency domain is expressed as
\begin{equation}
\widetilde{V}'_{\textrm{out}}[\omega]=U'(\omega)\widetilde{V}'_{\textrm{in}}[\omega]\label{eq:26}
\end{equation}
with the scattering matrix $U'(\omega)=\chi^{\textrm{T}}(P-i\omega I_{3})^{-1}\chi-I_{4}$.
Following the calculations similar to those in  Eqs.~(\ref{eq:14}-\ref{eq:18}), one obtains the expression for the output spectrum of VIS mode in the frequency domain as
\begin{equation}
\bar{S}'_{a,\textrm{out}}(\omega)=T'_{ac}(\omega)\left[S'_{c,\textrm{in}}(\omega)+\frac{1}{2}\right]+S'_{\textrm{add}}(\omega).\label{eq:27}
\end{equation}
Here, the conversion efficiency from the IR signal
to the VIS range is given by
\begin{eqnarray}
T'_{ac}(\omega) & = & \left|U'_{12}(\omega)\right|^{2},\label{eq:28}
\end{eqnarray}
where the related transmission amplitude $U'_{12}$ is
\begin{eqnarray}
U'_{12}(\omega) & = & \frac{2i\mathcal{G}_{a}G_{c}\sqrt{\kappa_{a}\kappa_{c}}}{\Lambda(\omega)}\label{30}
\end{eqnarray}
with $\Lambda(\omega)=(\Delta-\omega-i\kappa_{a})G_{c}^{2}+[\mathcal{G}_{a}^{2}-(\Delta-\omega-i\kappa_{a})(\omega_{b}-\omega-i\gamma_{B})](\omega_{c}-\omega-i\zeta).$ 
The added noise power $S'_{\text{add}}(\omega)$ is expressed as
\begin{equation}
S'_{\textrm{add}}(\omega)=\frac{\left|U'_{11}(\omega)\right|^{2}+\left|U'_{13}(\omega)\right|^{2}+\left|U'_{14}(\omega)\right|^2}{2},
\label{eq:29}
\end{equation}
where
\begin{eqnarray}
U'_{11}(\omega) & = &-1-\frac{2i\kappa_a[(G_c^2+(\omega_b-\omega-i\gamma_B)(\omega-\omega_c+i\zeta)]}{\Lambda(\omega)},\nonumber  \\
U'_{13}(\omega) & = &\frac{2i\mathcal{G}_{a}\sqrt{\gamma_B\kappa_c}(\omega-\omega_c+i\zeta)}{\Lambda(\omega)},\nonumber  \\
U'_{14}(\omega) & = &\frac{2i\sqrt{\gamma_B\kappa_a}\mathcal{G}_{a}G_c}{\Lambda(\omega)}.\label{eq:30}
\end{eqnarray}

When the effective detuning $\Delta$ is tuned to $\omega_{b}$ ($\Delta=\omega_{b}$) and the frequency of the IR mode is set to be resonant with the molecular vibrational mode ($\omega_{c}=\omega_{b})$, the conversion efficiency at the first anti-Stokes sideband of the VIS mode, $T_{ac}'^{\text{AS}}$, can be calculated as
\begin{eqnarray}
T_{ac}'^{\textrm{AS}} & \equiv & \left|U'_{12}(\omega=\omega_{b})\right|^{2}\nonumber  \\  
& = & \left|\frac{2\mathcal{G}_{a}G_{c}\sqrt{\kappa_{a}\kappa_{c}}}{G_{c}^{2}\kappa_{a}+\zeta(\gamma_{B}\kappa_{a}+\mathcal{G}_{a}^{2})}\right|^{2}.\label{eq:31}
\end{eqnarray}
Under experimentally relevant parameters, $\zeta\gamma_{B}\kappa_{a}\ll\{G_{c}^2\kappa_{a},~ \zeta\mathcal{G}_{a}\}$ is usually well satisfied. When $\zeta > 0$, the term $\zeta\gamma_{B}\kappa_{a}$ can be neglected, and the conversion efficiency at the first anti-Stokes sideband, $T_{ac}'^{\textrm{AS}}$, approaches its maximum value 1 at $G_{c}^{2}\kappa_{a}\simeq\ \zeta\mathcal{G}_{a}^{2}$. When  $\zeta < 0$, the cancellation between competing terms in the denominator instead renders the term $\zeta\gamma_{B}\kappa_{a}$ non-negligible in the calculation of the conversion efficiency $T_{ac}'^{\textrm{AS}}$.  The optimal conversion efficiency is achieved at
$G_{c}^{2}\kappa_{a}+\zeta(\gamma_{B}\kappa_{a} + \mathcal{G}_{a}^{2})\simeq 0 $. In the optimal case, the conversion efficiency becomes divergent at the first anti-Stokes sideband, indicating the breakdown of the RWA in this parameter regime.

\section{analysis}

\begin{figure}
\begin{centering}
\includegraphics[width=8cm,totalheight=4cm]{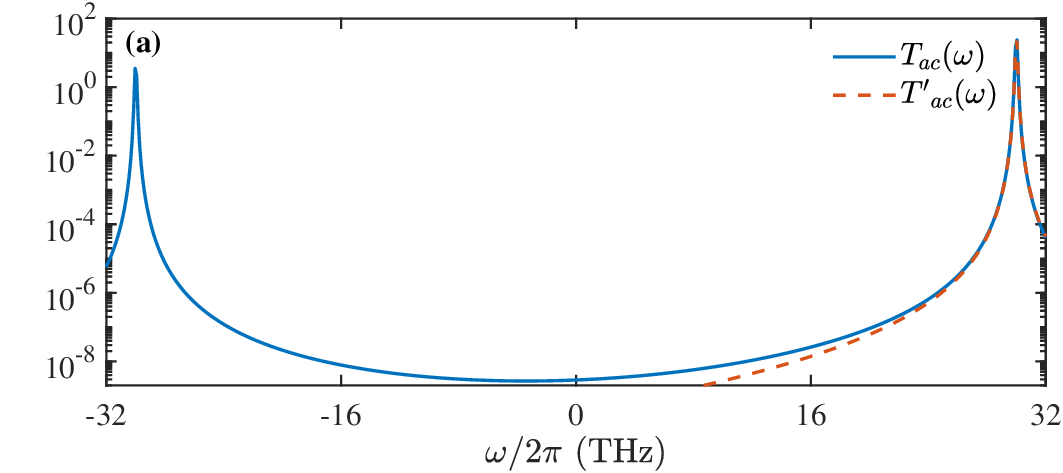}
\par\end{centering}
\begin{centering}
\includegraphics[width=8cm,totalheight=4cm]{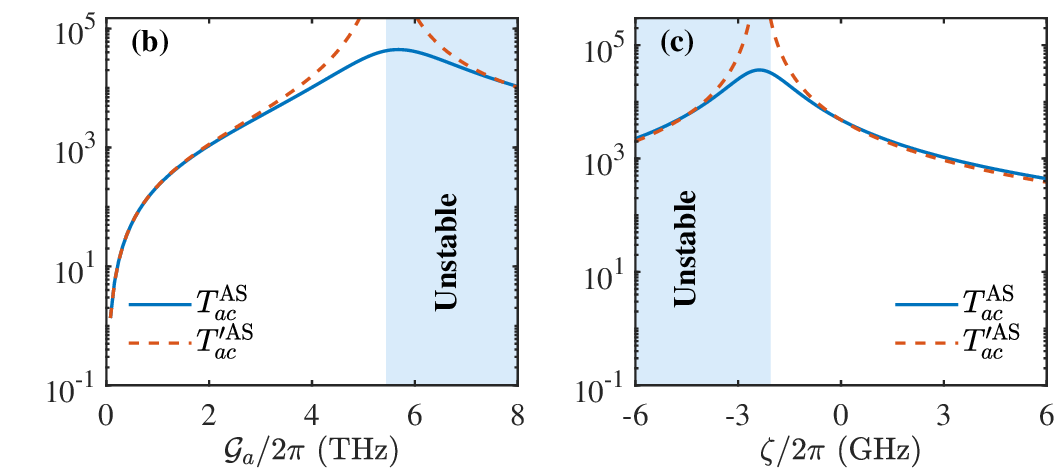}
\par\end{centering}
\centering{}\caption{(a) The conversion efficiencies $T_{ac}(\omega)$ and $T'_{ac}(\omega)$
 as functions of frequency $\omega$ for $\mathcal{G}_{a}/2\pi=800$ GHz at $\zeta/2\pi=0.03$ THz. (b,~c) The conversion efficiencies $T_{ac}^{\mathrm{AS}}(\omega=\omega_{b})$ and $T_{ac}'^{\mathrm{AS}}(\omega=\omega_{b})$ at the first anti-Stokes sideband  as functions of the enhanced collective molecular coupling strength $\mathcal{G}_{a}$ and the effective dissipation rate $\zeta$, respectively. In panel (b), $\zeta/2\pi=-2$ GHz. In panel (c), $\mathcal{G}_{a}/2\pi=5$ THz.  Other parameters are $\Delta/2\pi=\omega_{c}/2\pi=\omega_{b}/2\pi=30$ THz, $\kappa_{a}/2\pi=25$ THz, $\kappa_{c}/2\pi=3$ THz, $\gamma_{B}/2\pi=0.06$ THz, $N=10^{6}$, and $g_{c}/2\pi=0.05$ GHz. \label{fig:3}}
\end{figure}

To verify validity of the RWA, we quantitatively compare the conversion efficiencies obtained with and without the RWA. Figure~\ref{fig:3}(a) plots the conversion efficiencies $T_{ac}(\omega)$  and $T'_{ac}(\omega)$, derived from Eq.~(\ref{eq:19}) without the RWA and Eq.~(\ref{eq:25}) with the RWA, as functions of frequency. In the case without employing the RWA, two prominent peaks appear at the first Stokes and anti-Stokes sidebands (see the blue solid linear), with peak values of approximately 3.5 and 22. Under the RWA, it shows a single peak at the first anti-Stokes sideband (see the red dash linear), reaching a maximum of around 20. Evidently, the conversion efficiencies of the two cases are in agreement at the first anti-Stokes sideband.

 In addition, we examine how the conversion efficiency at the first anti-Stokes sideband varies with $\mathcal{G}_{a}$ and $\zeta$ under the RWA and non-RWA cases. Note that we focus on the conversion efficiency of the system in the stable regime. As shown in Fig.~\ref{fig:3}(b), within the regime of interest, as $\mathcal{G}_a$ increases, the conversion efficiencies $T^\text{AS}_{ac}$ and ${T}_{ac}'^{\text{AS}}$ at the first anti-Stokes sideband increase accordingly. The conversion efficiencies obtained under both frameworks exhibit a high degree of overlap across the  range of $\mathcal{G}_a/2\pi$ from 0 to 3 $\text{THz}$. Figure~\ref{fig:3}(c) depicts another case involving various $\zeta$. In the region $\zeta/2\pi \geq 0.5$ GHz, the two conversion efficiencies $T_{ac}(\omega)$ and ${T}_{ac}'(\omega)$ almost coincide, and both decrease as the effective dissipation rate $\zeta$ increases. These observations indicate that $T_{ac}^{AS}$ and $T_{ac}'^{\text{AS}}$ have a good qualitative and even quantitative consistency near the first anti-Stokes sideband ($\omega\sim\omega_b$), except within a certain parameter region. For example, around the regime at $\mathcal{G}_{a}/2\pi=5$ THz and $\zeta/2\pi=2$ GHz, the values of $T_{ac}^\text{AS}$ and ${T}_{ac}'^{\textrm{AS}}$ reach $3\times10^4$ and $3\times10^5$, respectively.
 \begin{figure}
\centering{}\includegraphics[width=9cm,totalheight=4cm]{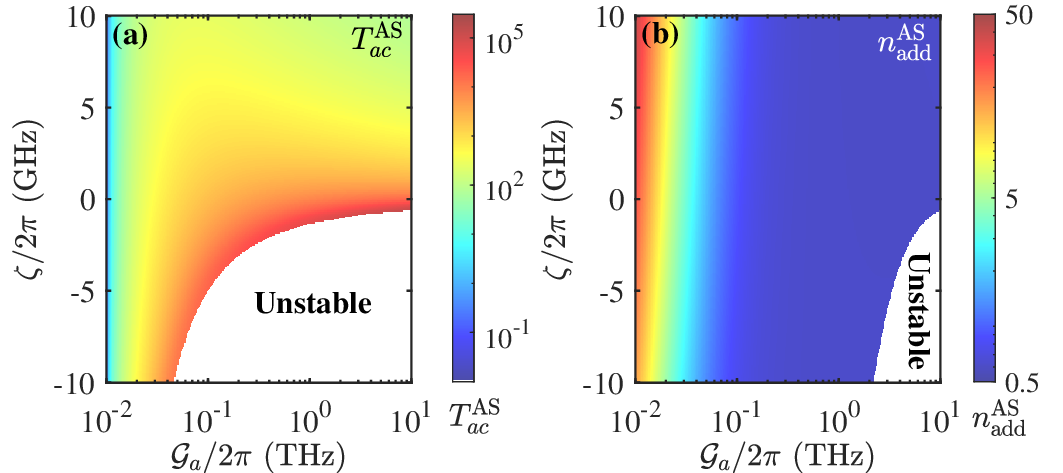}\caption{(a,~b) The conversion efficiency $T_{ac}^{\text{AS}}$
and the added noise $n_\text{add}^\text{AS}$ at the first anti-Stokes sideband ($\omega= \omega_{b}$) as functions of the enhanced optomechanical coupling strength $\mathcal{G}_{a}$ and the effective dissipation rate $\zeta$. The parameters are $g_{c}/2\pi=0.05$ GHz, $\Delta/2\pi=\omega_{c}/2\pi=\omega_{b}/2\pi=30$ THz, $\kappa_{a}/2\pi=25$
THz, $\kappa_{c}/2\pi=3$ THz, $\gamma_{B}/2\pi=0.06$ THz, and $N=10^{6}$.\label{fig:4}}
\end{figure}
\begin{figure}
\centering{}\includegraphics[width=8cm,totalheight=4cm]{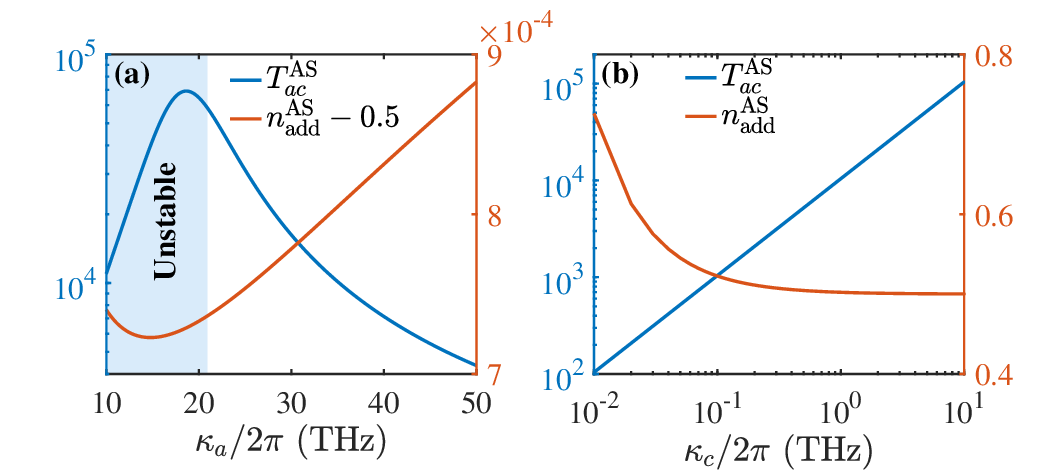}\caption{The conversion efficiency $T_{ac}^{\text{AS}}$ and the added noise $n_\text{add}^\text{AS}$ at the first anti-Stokes sideband ($\omega= \omega_{b}$) as functions of (a) the decay rate of the VIS mode $\kappa_a$ and (b) the decay rate of the IR mode $\kappa_c$. In panel~(a), the parameter $\kappa_c/2\pi$ is set to $\kappa_c/2\pi=3$ THz. In panel~(b), the parameter $\kappa_a/2\pi$ is set to $\kappa_a/2\pi=25$ THz. Other parameters are $\mathcal{G}_{a}/2\pi=5$ THz, $\zeta/2\pi=-2$ GHz, $g_{c}/2\pi=0.05$ GHz, $\Delta/2\pi=\omega_{c}/2\pi=\omega_{b}/2\pi=30$ THz, $\gamma_{B}/2\pi=0.06$ THz, and $N=10^{6}$.}\label{fig:5}
\end{figure}Accounting for the difference, we select the conversion efficiency $T_{ac}^\text{AS}$ without the RWA for further analysis to ensure reliability. 

As presented in Fig.~\ref{fig:4}(a), the conversion efficiency $T_{ac}^\text{AS}$ can be optimized to $3\times10^4$ or more by cooperatively decreasing the parameter $\zeta$ and increasing the coupling strength $\mathcal{G}_{a}$.  The increase in the coupling strength $\mathcal{G}_{a}$ and the gain $g$ inevitably results in a change of the added noise power $S_{\text{add}}$. To evaluate its impact on the upconversion signal, we introduce the added noise defined as~\cite{PhysRevLett.120.023601,hill2012coherent}
\begin{eqnarray}
n_{\textrm{add}}(\omega)=\frac{S_{\textrm{add}}(\omega)}{{T}^\text{AS}_{ac}}.\label{eq:34}
\end{eqnarray}
 Figure~\ref{fig:4}(b) shows the added noise at the first anti-Stokes sideband, $n_{\text{add}}^{\text{AS}}$ ($\equiv n_{\text{add}}(\omega=\omega_b)$), as functions of parameters  $\mathcal{G}_{a}$ and $\zeta$. By analyzing the individual effect of the parameters $\mathcal{G}_{a}$ and $\zeta$ on the added noise, we find that when $\mathcal{G}_{a}$ is close to zero, the added noise exceeds 1. With increasing $\mathcal{G}_{a}$, the added noise initially decreases rapidly and eventually reaches a saturation value of approximately 0.5. Crucially, it can be seen from Figs.~\ref{fig:4}(a) and~\ref{fig:4}(b) that in the parameter regime where the added noise approaches 0.5, the conversion efficiency is significantly larger than 1. The results mean that introducing gain into IR cavity achieves efficient signal conversion with significant amplification while keeping low added noise.

In the following, we discuss the dependence of the conversion efficiency ${T}_{ac}^{\textrm{AS}}$ and the added noise $n_\text{add}^\text{AS}$ on other parameters such as the decay rates of the VIS and  IR modes, $\kappa_{a}$ and $\kappa_c$. As illustrated in Fig.~\ref{fig:5}(a), within the stable region, increasing the parameter $\kappa_a$ results in a marked decrease in the conversion efficiency $T_\text{add}^\text{AS}$. Meanwhile, the added noise $n_\text{add}^\text{AS}$ grows slightly.  When varying the parameter $\kappa_{c}$, an entirely different trend is shown in Fig.~\ref{fig:5}(b). With increasing parameter $\kappa_{c}$, the conversion efficiency $T_\text{ac}^\text{AS}$ is significantly enhanced and the added noise falls rapidly to about 0.5. These results indicates that the system with gain can obtain a high-efficiency frequency conversion (e.g. $10^5$) at low added noise (e.g. 0.5) by taking appropriate values of $\kappa_a$ and $\kappa_c$.

\section{concussion}

In summary, we have proposed an alternative scheme to enhance the frequency upconversion of infrared signals under red-detuned conditions in the molecular cavity optomechaincal system. The introduction of the gain in the IR cavity significantly boosts the conversion efficiency from the IR to the VIS range while keeping the added noise close to 0.5.  The high conversion efficiency achieved in the system with gain facilitates sensitive and precise detection of IR signals. Our work provides a theoretical support for experimental developments in IR signal upconversion.

\section*{Acknowledgments}

This work is supported by the Quantum Science and Technology-National Science and Technology Major Project (Grant No. 2023ZD0300700), the National Natural Science Foundation of China (Grants No. 12574387, No. 12405011, and No. 12547103), and the Natural Science Foundation of Hainan Province (Grants No. 125QN210 and No. 125RC631).


%

\end{document}